# Criterion of stability of the superconducting state

Iogann Tolbatov

*Physics and Engineering Department, Kuban State University, Krasnodar, Russia*

(talbot1038@mail.ru)

In this paper, we propose to draw attention to the stability criterion of the superconductor current state. We use for this purpose the rough systems mathematical apparatus allowing us to relate the desired criterion with the dielectric permittivity of the matter and to identify the type of all possible phonons trajectories in its superconducting state. The state of superconductivity in the matter can be explained only by the phonons behavior peculiarity. And on the basis of the above-mentioned assumption, the corresponding mathematical model is constructed.

The study of superconductivity continues already for more than 50 years. Cooper's basic work [1] in 1956 has explained the mechanism of superconductivity on the basis of the assumption about the coupling of electrons in Cooper pairs. Then followed the microscopic theory of superconductivity of Bardeen-Cooper-Schrieffer [2]. Luttinger [3] tried to explain the phenomenon of high-temperature superconductivity (HTSC) on the basis of the separation of spin and charge, and regular researches of Anderson [4], [5] developing ideas about Luttinger liquid as about the essence of electronic systems in HTSC; works of Laughlin [6], in which the fractional statistics for the description of low-energy excitations in HTSC systems was used, anyhow described the condition of low- or high- temperature superconductivity, but the unified theory of superconductivity was not been created.

In our research other model of this phenomenon is offered, it describes the most general principles of low- and high- temperature superconductivity. Its methodological basis became the mathematical apparatus of rough systems described by L. S. Pontryagin [7].

We consider the system of the charged particles, on which external forces do not influence, that is: the particles either rest or move without acceleration (there is no second derivative on time):

$$\frac{dx}{dt} = P(x,y); \quad \frac{dy}{dt} = Q(x,y); \tag{1}$$

where $x = x(t)$, $y = y(t)$ – particles trajectories, $P(x,y)$, $Q(x,y)$ – analytical functions.

Dependence of particle trajectories from time $t$ indicates the presence of the so-called "cycle without contact" because in the absence of external (and dissipative) forces, the total energy $E$ is constant, and the total work of force acting on a particle is: $\oint_L A dl = 0$, and the trajectory $L$ is closed. In this case, there exists a domain $G$ bounded by simple closed curve $g$ with a continuously rotating tangent. Outside of $G$, the system has no "cycle without contact".

Now we consider alongside with the system (1) the changed system:

$$\frac{dx}{dt} = P(x,y) + p(x,y); \quad \frac{dy}{dt} = Q(x,y) + q(x,y); \tag{2}$$

where $p(x,y)$ and $q(x,y)$ are the analytical functions, for which the following conditions are true:

$\forall \delta > 0; \; \forall \varepsilon > 0$

$\exists \forall p(x,y): \; |p(x,y)| < \varepsilon; \; |p'_x(x,y)| < \varepsilon; \; |p'_y(x,y)| < \varepsilon;$

$\exists \forall q(x,y): \; |q(x,y)| < \varepsilon; \; |q'_x(x,y)| < \varepsilon; \; |q'_y(x,y)| < \varepsilon.$

There is a mutually unique and mutually continuous transformation of $K$ of $G$ domain into itself in which: 1) relevant to each other points are at a distance of less than $\delta$, 2) the points lying on the same trajectory of system (1) correspond to the points lying on the same trajectory of system (2), and vice versa. Thus, in G, the system (1) is a rough system.

Since we are using the apparatus of rough systems (1), we shall pay attention to their three properties:

I. If the system (1) is rough in G, then in G, the system (1) can only have such an equilibrium position, for which the real parts of the roots of the characteristic equation are different from zero.

Consequently, the system (1) in G can not have equilibrium $x = x_0$ and $y = y_0$, for which the following is true:

a) $\Delta = 0$, (3)

b) $\Delta > 0$, $\sigma = -(P'_x(x_0, y_0) + Q'_y(x_0, y_0)) = 0$, (4)

where $\Delta = \begin{vmatrix} P'_x(x_0, y_0) & P'_y(x_0, y_0) \\ Q'_x(x_0, y_0) & Q'_y(x_0, y_0) \end{vmatrix}$.

II. If the system (1) is rough in the field of G, then in G, the system (1) can have only such periodic motion, for which the characteristic parameter is not zero. In other words: in G, the system (1) does not have periodic motion $x = \varphi(t)$, $y = \psi(t)$, $[\varphi(t + \tau) = \varphi(t); \psi(t + \tau) = \psi(t)]$, for which

$$h = \frac{1}{\tau} \int_0^\tau [P'_x(\varphi, \psi) + Q'_y(\varphi, \psi)] dt = 0.$$ (5)

III. If the system (1) is rough in the field of G, then in G the system (1) can only have such a separatrix that does not go from saddle into saddle.

Thus, the rough system (1) within the "cycle without contact" $g$ has the following properties:

a) only such states of equilibrium, for which $\Delta \neq 0$; and if $\Delta > 0$, then $\sigma \neq 0$,

b) only such periodic trajectories, for which $h \neq 0$,

c) only such separatrices, which do not go from saddle into saddle.

We find the stability criterion for a system of electrons. The potential of electron-electron interaction has the property of asymptotic freedom, i. e. the total effect of the Coulomb potential and the potential of deformation of the crystal lattice are such that the electrons of the Cooper pairs move quasifree. Therefore, they should be regarded as free electrons, then the equation of electron motion in electric field of an external monochromatic electromagnetic wave has the form [8]:

$$m\ddot{x} = eE_0 e^{-i\omega t}. \tag{6}$$

From the equation (6), we obtain the law of the frequency dispersion of the electron system permittivity in a standard way:

$$\varepsilon(\omega) = 1 - \frac{\omega_p^2}{\omega^2}, \tag{7}$$

where $\omega_p$ is the frequency of plasma oscillations.

Expressing from (7) the oscillation frequency $\omega = \frac{\omega_p}{\sqrt{1-\varepsilon}}$ through the dielectric constant and substituting the latter in (6), we find the equations of motion of two quasifree electrons forming a Cooper pair as the functions of the coordinates of the two deformation potentials, which determine the local values of the permittivities $\varepsilon_1(x,y)$ and $\varepsilon_2(x,y)$ corresponding to the electron localization coordinates $x$ and $y$.

$$\frac{dx}{dt} = \frac{ieE_0}{m\omega_p}\sqrt{1-\varepsilon_1(x,y)}\exp\left(-\frac{i\omega_p t}{\sqrt{1-\varepsilon_1(x,y)}}\right),$$

$$\frac{dy}{dt} = \frac{ieE_0}{m\omega_p}\sqrt{1-\varepsilon_2(x,y)}\exp\left(-\frac{i\omega_p t}{\sqrt{1-\varepsilon_2(x,y)}}\right).$$

Then we find $P'_x$, $P'_y$, $Q'_x$, $Q'_y$

$$P'_x = \frac{C\varepsilon'_{1x}}{\sqrt{1-\varepsilon_1}}\exp\left(-\frac{i\omega_p t}{\sqrt{1-\varepsilon_1}}\right)\left(\frac{i\omega_p t}{\sqrt{1-\varepsilon_1}}+1\right),$$

$$P'_y = \frac{C\varepsilon'_{1y}}{\sqrt{1-\varepsilon_1}}\exp\left(-\frac{i\omega_p t}{\sqrt{1-\varepsilon_1}}\right)\left(\frac{i\omega_p t}{\sqrt{1-\varepsilon_1}}+1\right),$$

$$Q'_x = \frac{C\varepsilon'_{2x}}{\sqrt{1-\varepsilon_2}}\exp\left(-\frac{i\omega_p t}{\sqrt{1-\varepsilon_2}}\right)\left(\frac{i\omega_p t}{\sqrt{1-\varepsilon_2}}+1\right),$$

$$Q'_y = \frac{C\varepsilon'_{2y}}{\sqrt{1-\varepsilon_2}}\exp\left(-\frac{i\omega_p t}{\sqrt{1-\varepsilon_2}}\right)\left(\frac{i\omega_p t}{\sqrt{1-\varepsilon_2}}+1\right),$$

where $C = \dfrac{ieE_0}{2m\omega_p}$. Calculating the discriminant $\Delta = \begin{vmatrix} P'_x(x_0, y_0) & P'_y(x_0, y_0) \\ Q'_x(x_0, y_0) & Q'_y(x_0, y_0) \end{vmatrix}$, we find

$$\Delta = ZD, \qquad (8)$$

where

$$Z = C^2 \exp\left[-i\omega_p t \left(\frac{1}{\sqrt{1-\varepsilon_1}} + \frac{1}{\sqrt{1-\varepsilon_2}}\right)\right] \frac{1}{\sqrt{(1-\varepsilon_1)(1-\varepsilon_2)}} \left(1 + \frac{i\omega_p t}{\sqrt{1-\varepsilon_1}}\right)\left(1 + \frac{i\omega_p t}{\sqrt{1-\varepsilon_2}}\right)$$

and $D = \begin{vmatrix} \varepsilon'_{1x} & \varepsilon'_{1y} \\ \varepsilon'_{2y} & \varepsilon'_{2y} \end{vmatrix}$.

The condition (3) corresponds to the determinant $D$ vanishing. The multiplier $Z$ can not be zero, as it follows from its explicit form.

Now we verify the condition (4). In the case of $\Delta > 0$, multipliers $Z$ and $D$ have the same sign. The parameter $\sigma$ becomes zero only if $P'_x = -Q'_y$ or

$$C \frac{\varepsilon'_{1x}}{\sqrt{1-\varepsilon_1}} \exp\left(-\frac{i\omega_p t}{\sqrt{1-\varepsilon_1}}\right)\left(\frac{i\omega_p t}{\sqrt{1-\varepsilon_1}} + 1\right) = -C \frac{\varepsilon'_{2y}}{\sqrt{1-\varepsilon_2}} \exp\left(-\frac{i\omega_p t}{\sqrt{1-\varepsilon_2}}\right)\left(\frac{i\omega_p t}{\sqrt{1-\varepsilon_2}} + 1\right).$$

From the last equality, it follows that

$$\frac{\varepsilon'_{1x}}{\varepsilon'_{2y}} = \frac{H(\varepsilon_2)}{H(\varepsilon_1)}, \qquad (9)$$

where $H(\varepsilon_1)$ and $H(\varepsilon_2)$ denote $\dfrac{1}{\sqrt{1-\varepsilon_1}} \exp\left(-\dfrac{i\omega_p t}{\sqrt{1-\varepsilon_1}}\right)\left(\dfrac{i\omega_p t}{\sqrt{1-\varepsilon_1}} + 1\right)$ and

$\dfrac{1}{\sqrt{1-\varepsilon_2}} \exp\left(-\dfrac{i\omega_p t}{\sqrt{1-\varepsilon_2}}\right)\left(\dfrac{i\omega_p t}{\sqrt{1-\varepsilon_2}} + 1\right)$ respectively.

So, the conditions stated in (4) are equivalent to the following:

1) $\begin{cases} D > 0 \\ H(\varepsilon_1) > 0 \\ H(\varepsilon_2) > 0 \end{cases}$, or 2) $\begin{cases} D < 0 \\ H(\varepsilon_1) < 0 \\ H(\varepsilon_2) < 0 \end{cases}$, or 3) $\begin{cases} D < 0 \\ H(\varepsilon_1) < 0 \\ H(\varepsilon_2) > 0 \end{cases}$, or 4) $\begin{cases} D < 0 \\ H(\varepsilon_1) > 0 \\ H(\varepsilon_2) < 0 \end{cases}$.

This means:

1) if $\varepsilon'_{1x}\varepsilon'_{2y} > \varepsilon'_{1y}\varepsilon'_{2x}$ : $\dfrac{\varepsilon'_{1x}}{\varepsilon'_{2y}} > 0 \Leftrightarrow \begin{cases} \varepsilon_1 > 1 + t^2\omega_p^2 \\ \varepsilon_2 > 1 + t^2\omega_p^2 \end{cases}$,

2) if $\varepsilon'_{1x}\varepsilon'_{2y} > \varepsilon'_{1y}\varepsilon'_{2x}$: $\dfrac{\varepsilon'_{1x}}{\varepsilon'_{2y}} > 0 \Leftrightarrow \begin{cases} \varepsilon_1 < 1+t^2\omega_p^2 \\ \varepsilon_2 < 1+t^2\omega_p^2 \end{cases}$,

3) if $\varepsilon'_{1x}\varepsilon'_{2y} < \varepsilon'_{1y}\varepsilon'_{2x}$: $\dfrac{\varepsilon'_{1x}}{\varepsilon'_{2y}} < 0 \Leftrightarrow \begin{cases} \varepsilon_1 < 1+t^2\omega_p^2 \\ \varepsilon_2 > 1+t^2\omega_p^2 \end{cases}$,

4) if $\varepsilon'_{1x}\varepsilon'_{2y} < \varepsilon'_{1y}\varepsilon'_{2x}$: $\dfrac{\varepsilon'_{1x}}{\varepsilon'_{2y}} < 0 \Leftrightarrow \begin{cases} \varepsilon_1 > 1+t^2\omega_p^2 \\ \varepsilon_2 < 1+t^2\omega_p^2 \end{cases}$.

Now we consider the case of system stability, the criterion of which is the condition $\Delta < 0$, in which the factors $Z$ and $D$ have different signs. If $D < 0$ and $\varepsilon'_{1x}\varepsilon'_{2y} < \varepsilon'_{1y}\varepsilon'_{2x}$, the following two cases are possible:

1) $\begin{aligned} H(\varepsilon_1) &> 0 \Leftrightarrow \varepsilon_1 > 1+t^2\omega_p^2 \\ H(\varepsilon_2) &> 0 \quad \varepsilon_2 < 1+t^2\omega_p^2 \end{aligned}$;

2) $\begin{aligned} H(\varepsilon_1) &< 0 \Leftrightarrow \varepsilon_1 < 1+t^2\omega_p^2 \\ H(\varepsilon_2) &< 0 \quad \varepsilon_2 < 1+t^2\omega_p^2 \end{aligned}$.

If $D > 0$ and $\varepsilon'_{1x}\varepsilon'_{2y} > \varepsilon'_{1y}\varepsilon'_{2x}$, the following two cases are possible:

3) $\begin{aligned} H(\varepsilon_1) &> 0 \Leftrightarrow \varepsilon_1 > 1+t^2\omega_p^2 \\ H(\varepsilon_2) &< 0 \quad \varepsilon_2 < 1+t^2\omega_p^2 \end{aligned}$;

4) $\begin{aligned} H(\varepsilon_1) &< 0 \Leftrightarrow \varepsilon_1 < 1+t^2\omega_p^2 \\ H(\varepsilon_2) &> 0 \quad \varepsilon_2 > 1+t^2\omega_p^2 \end{aligned}$.

Summarizing the results of the above analysis, we conclude that the system is unstable, when $D = 0$ or equality is true (9).

The system is stable in following cases:

$\varepsilon'_{1x}\varepsilon'_{2y} > \varepsilon'_{1y}\varepsilon'_{2x}$, $\varepsilon_{1,2} < 1$ $\qquad \begin{cases} \varepsilon_1 > 1+t^2\omega_p^2 \\ \varepsilon_2 > 1+t^2\omega_p^2 \end{cases}; \dfrac{\varepsilon'_{1x}}{\varepsilon'_{2y}} > 0,$

$\begin{cases} \varepsilon_1 < 1+t^2\omega_p^2 \\ \varepsilon_2 < 1+t^2\omega_p^2 \end{cases}; \dfrac{\varepsilon'_{1x}}{\varepsilon'_{2y}} > 0$

$\begin{cases} \varepsilon_1 > 1+t^2\omega_p^2 \\ \varepsilon_2 < 1+t^2\omega_p^2 \end{cases}$

$\begin{cases} \varepsilon_1 < 1+t^2\omega_p^2 \\ \varepsilon_2 > 1+t^2\omega_p^2 \end{cases}$

$$\varepsilon'_{1x}\varepsilon'_{2y} < \varepsilon'_{1y}\varepsilon'_{2x}, \quad \varepsilon_{1,2} < 1 \quad \begin{cases} \varepsilon_1 < 1+t^2\omega_p^2 \\ \varepsilon_2 > 1+t^2\omega_p^2 \end{cases}; \frac{\varepsilon'_{1x}}{\varepsilon'_{2y}} < 0,$$

$$\begin{cases} \varepsilon_1 > 1+t^2\omega_p^2 \\ \varepsilon_2 < 1+t^2\omega_p^2 \end{cases}; \frac{\varepsilon'_{1x}}{\varepsilon'_{2y}} < 0$$

$$\begin{cases} \varepsilon_1 > 1+t^2\omega_p^2 \\ \varepsilon_2 > 1+t^2\omega_p^2 \end{cases}$$

$$\begin{cases} \varepsilon_1 < 1+t^2\omega_p^2 \\ \varepsilon_2 < 1+t^2\omega_p^2 \end{cases}$$

These conditions must be fulfilled for any $t$. The stability criterion can be formulated as follows:

a) $\varepsilon'_{1x}\varepsilon'_{2y} > \varepsilon'_{1y}\varepsilon'_{2x}$ $\begin{cases} \varepsilon_1 < 1+t^2\omega_p^2 \\ \varepsilon_2 < 1+t^2\omega_p^2 \end{cases}; \frac{\varepsilon'_{1x}}{\varepsilon'_{2y}} > 0$;

b) $\varepsilon'_{1x}\varepsilon'_{2y} < \varepsilon'_{1y}\varepsilon'_{2x}$ $\begin{cases} \varepsilon_1 < 1+t^2\omega_p^2 \\ \varepsilon_2 < 1+t^2\omega_p^2 \end{cases}; \varepsilon_{1,2} < 1$.

Thus, the criterion of stability of the charged particles system has the form:

$$\begin{bmatrix} a)\varepsilon'_{1x}\varepsilon'_{2y} > \varepsilon'_{1y}\varepsilon'_{2x}; \dfrac{\varepsilon'_{1x}}{\varepsilon'_{2y}} > 0; \varepsilon_{1,2} < 1 \\ \textit{б})\varepsilon'_{1x}\varepsilon'_{2y} < \varepsilon'_{1y}\varepsilon'_{2x}; \varepsilon_{1,2} < 1 \end{bmatrix}$$

The system of charged particles is unstable if $\varepsilon'_{1x}\varepsilon'_{2y} = \varepsilon'_{1y}\varepsilon'_{2x}$.

Now we calculate the periodic trajectories $h$, at which the system (1) moves within the "cycle without contact" $g$. Using (5) and the resulted above values of derivatives $P'_x$ and $Q'_y$, we find that $h = h_1 + h_2$, where

$$h_1 = -\frac{C\varepsilon'_{1x}}{\tau}\left[\frac{2}{i\omega_p}\left(\exp\left(-\frac{i\omega_p\tau}{\sqrt{1-\varepsilon_1}}\right)-1\right) + \frac{\tau}{\sqrt{1-\varepsilon_1}}\exp\left(-\frac{i\omega_p\tau}{\sqrt{1-\varepsilon_1}}\right)\right],$$

$$h_2 = -\frac{C\varepsilon'_{2y}}{\tau}\left[\frac{2}{i\omega_p}\left(\exp\left(-\frac{i\omega_p\tau}{\sqrt{1-\varepsilon_2}}\right)-1\right) + \frac{\tau}{\sqrt{1-\varepsilon_2}}\exp\left(-\frac{i\omega_p\tau}{\sqrt{1-\varepsilon_2}}\right)\right].$$

Using the relation (9) and our analysis, we conclude that $h$ does not turn into zero.

We have examined the movement of monochromatic free electrons, which, as we believe, can describe the state of superconductivity in the matter. We have shown that in the absence of dissipative forces in the *G* domain limited by the simple closed curve *g* with a continuously rotating tangent, there exists the "cycle without contact" for the dynamic system. The research results show that the dynamic system is rough and possesses the listed properties: stability and the closed trajectories.

Thus, we treat the stability in article in a broad sense: and as the absence (filling up) of dissipation finally, and as the necessary condition of superconductivity. Movement of conductivity electrons is considered by us as the function of the phonons coordinates defined by the dielectric permittivities $\varepsilon_1(x,y)$ and $\varepsilon_2(x,y)$ caused by them. The trajectory of a phonon is the attractor of electron gas movement, therefore, we have described the current of superconductivity through the oscillations of the crystal lattice site.

Study of the presented rough system has allowed:

1) to identify the desired stability criterion:

$$\begin{bmatrix} a)\varepsilon'_{1x}\varepsilon'_{2y} > \varepsilon'_{1y}\varepsilon'_{2x}; \ \dfrac{\varepsilon'_{1x}}{\varepsilon'_{2y}} > 0; \ \varepsilon_{1,2} < 1 \\ б)\varepsilon'_{1x}\varepsilon'_{2y} < \varepsilon'_{1y}\varepsilon'_{2x}; \ \varepsilon_{1,2} < 1 \end{bmatrix}, \qquad (10)$$

2) to determine the form of all possible trajectories with components defined by dynamic equations,

3) to conclude that the dynamic equations system has in *G* only the separatrices not going from saddle into saddle.

In dynamic system, for which conditions of continuity in *G* are true, the set of special trajectories (belonging to the first three types defined by Bendixon [9]) divide the region *G* into a finite number of connected components filled by ordinary trajectories (belonging to the fourth and fifth types by Bendixon [9]).

These components are divided into two classes: the class of the components adjacent to the "cycle without contact" *g*, and the class of internal components. In each internal component, any trajectory is positively or negatively stable in the

Lyapunov stability sense. Every such a component has within its boundaries one positively Lyapunov stable special trajectory representing "an element of attraction" or "drain", and one negatively Lyapunov stable special trajectory being "an element of pushing" or "source".

In each component adjacent to the "cycle without contact", any trajectory is positively Lyapunov stable, and each such component has within the boundaries one positive Lyapunov stable special trajectory – "drain".

From the study, it can be assumed for the physics of superconductivity: the state of superconductivity in the matter can be explained only by the phonons behavior peculiarity. The mathematical expression of the superconductivity model constructed by us uses the apparatus of rough systems.